\documentstyle[epsf,prd,preprint,aps]{revtex}
\begin{document}
\draft
\title{Instanton content of finite temperature QCD matter}
\author{M.-C. Chu$^{1}$ and S. Schramm${}^{2,3}$}
\address{$^{1}$ W.~K.~Kellogg Radiation Laboratory, 106-38, \\
California Institute of Technology, Pasadena, California 91125}
\address{$^{2}$ Nuclear Theory Center, Indiana University,
Bloomington, Indiana 47408}
\address{$^{3}$ GSI, 64220 Darmstadt, Germany}
\date{\today}
\maketitle
\tighten

\begin{abstract}
We investigate the temperature dependence of the instanton content of
gluon fields and their contribution to quark correlation using
quenched lattice QCD and the cooling method. We found a suppression
of the topological susceptibility at finite temperature, agreeing
with the PCAC expectation at low temperature and enhanced suppression
for temperatures at and above the deconfinement transition. For
temperatures up to about 334~MeV, the topological charge correlation
agrees well with a single instanton profile, though the size
parameter seems to change across the phase transition. The screening
wave functions for the pion and the rho become slightly more compact
at higher temperatures. Lattice cooling shows no contribution from
the instantons to the screening wave functions at temperatures above
the phase transition.
\end{abstract}

\narrowtext
\section{Introduction}
\label{secintro}

Lattice QCD calculations \cite{lat} have suggested the existence of a
deconfinement phase transition at a critical temperature $T_c$, at
which quark and gluons are deconfined and form a new phase of matter,
the so-called {\it quark-gluon plasma}. This possibility stimulated
numerous theoretical and phenomenological works on the subject, many
of which focus on the possible experimental signatures of the
quark-gluon plasma \cite{sig}. Whereas these works are extremely
useful, in view of the experiments at AGS, CERN, and RHIC,
theoretical studies of the properties of finite temperature QCD
matter should also be important. Lattice gauge theory provides a
useful tool to calculate various hadronic matrix elements from first
principle, and it therefore enables us to study static finite
temperature properties of QCD matter.

A particularly important and interesting question regarding QCD
matter at finite temperature is what happens to the hadrons. Most of
the phenomenological works on quark-gluon plasma signatures to date
are based on a picture of weakly interacting quarks and gluons at
temperatures above $T_c$. However, such a simple picture is not
completely supported by lattice calculations. For example, lattice
calculations of the so-called screening wave functions indicated
strong correlations among the quarks in hadronic channels
\cite{scor}. However, it is still an open question whether the
screening wave function results have any direct bearing on the
structure of hadrons at finite temperature.

In lack of a method to calculate the finite-temperature hadron wave
functions directly, we turn our attention to a class of gauge
configurations known to have important effects on the structure and
masses of light hadrons at zero temperature \cite{chn}, namely,
large, isolated instantons. The primary goal of this paper is to
study the instanton content of QCD matter and their contribution to
quark correlations at finite temperature.

It is well known that instantons break the $U_A(1)$ chiral symmetry,
resulting in the famous axial anomalies. At high temperature, Debye
screening gives rise to an electric mass for the gluons and thereby
suppresses instanton amplitudes; the $U_A(1)$ symmetry should
therefore be restored. Whether the $U_A(1)$ restoration at $T_U$
occurs at the same temperature as the $SU(N_f)_A$ restoration at
$T_c$ bears important consequences to the hadronic physics at finite
temperature: if $T_U=T_c$, the phase transition is of first order
\cite{wilpis} and there are drastic changes to hadrons at $T_c$; if,
on the other hand, $T_U>T_c$, the phase transition is of second
order. The two scenarios give rise to very different hadron physics
at temperatures around $T_c$ \cite{shur1}. One of our goals is to
locate $T_U$ via lattice calculation of the topological
susceptibility, which is a measure of the instanton content, and
thereby differentiate the two scenarios.

There have been calculations of the temperature dependence of the
topological susceptibility in pure SU(2) \cite{tep,dig}, pure SU(3)
\cite{htw}, as well as unquenched SU(3) with two light quark flavors
\cite{got}. However, these authors vary the temperature by adjusting
the coupling constant, making it tricky to compare the susceptibility
at different temperatures because of the different lattice length
scales. At larger $\beta$ finite size effects may plague the
calculations using small (e.g., $8^3 \times 4$) lattices. There is
also some confusion on the behavior of the susceptibility in the
SU(2) theory. Whereas Teper {\it et~al.}\ observed a sudden
suppression of the topological susceptibility at the deconfinement
temperature for both SU(2) and SU(3) fields \cite{tep,htw},
Di~Giacomo {\it et~al.}\ reported that the cooling method gave
ambiguous results for SU(2) theory, and the susceptibility determined
by their field theoretical method stays almost constant across the
phase transition \cite{dig}.

In this paper, we present calculations of the temperature dependence
of the instanton content of quenched SU(3) fields using the cooling
method. We vary the temperature by changing the number of time slices
and report results for $N_t=4$, 6, 8, 10, 12, 14, 16, $N_s=16$, and
$\beta=6$ lattices. We did not observe the ambiguity reported for the
SU(2) calculation, and the topological susceptibility is very stable
with respect to the number of cooling steps. We found a suppression
of the topological susceptibility starting at temperatures below
$T_c$. The topological charge correlation function agrees well with
the continuum instanton profile, although there seems to be a change
in the instanton size parameter across the phase transition. For
three temperatures around $T_c$, we also calculate the
spatially-propagated wave functions, or the so-called screening wave
functions, for the pion and the rho with and without cooling. We
found that while the uncooled screening wave functions are only
mildly sensitive to temperature, in agreement with previous
calculations \cite{scor}, the cooled wave functions for $T>200$~MeV
are indistinguishable from the free ones. We thus conclude that at
such high temperatures, instantons do not contribute to the screening
wave functions.

We first describe in Sec.~II the set of observables we calculated. A
brief outline of the cooling method is presented in Sec.~III, and our
main results are summarized in Sec.~IV. We then close with some
discussions in Sec.~V.

\section{Observables}
\label{secbasic}

The instanton content of the gauge fields can be monitored by the the
topological charge density, which can be defined on the lattice as
\begin{equation}
Q(x_n)=-{1\over 32 \pi^2} \epsilon _{\alpha \beta \gamma \delta}
{\rm Re~Tr} \left[ U_{\alpha \beta} (x_n) U_{\gamma \delta} (x_n)
\right] \;,
\label{qn}
\end{equation}
where $U_{\alpha \beta}$ is the product of the link variables around
a plaquette in the $\alpha - \beta$ plane, after some cooling steps
were applied to remove lattice artifacts associated with
discretization \cite{htw}. The topological susceptibility is then
given as the fluctuations of the topological charge:
\begin{equation}
\chi _t \equiv {1 \over N_t N_s^3 a^4}\langle \left( \sum _n Q(x_n)
\right)^2 \rangle \;,
\label{chi}
\end{equation}
where $\langle\ldots\rangle$ indicates configuration averaging, and
$N_t$ and $N_s$ are the number of sites in the temporal and spatial
direction. The lattice spacing $a$ changes as a function of the
lattice inverse coupling $\beta$, which makes it tricky to compare
$\chi _t$ calculated with different $\beta$, especially when $\beta$
is not large enough to be in the asymptotic scaling regime. To study
the temperature dependence of the topological susceptibility, we
chose to change the temperature by varying the number of time slices
\begin{equation}
T={1\over N_t a} \;,
\label{temp}\end{equation}
but keeping fixed $\beta$ and hence $a$. This way the uncertainty in
$a$ enters only as an overall constant factor which does not affect
the shape of the $\chi _t$ vs.\ $T$ curve.

In order to look into the details of the topological charge
distribution, we also calculated the topological charge density
correlation function
\cite{chn}
\begin{equation}
C_Q(x)=\langle \sum _y Q(y) Q(x+y) \rangle / \langle \sum _y Q^2(y)
\rangle
\;.
\label{cprofile}
\end{equation}
One can compare this correlation function to a convolution of an
isolated instanton topological charge density
\begin{equation}
Q_\rho (x)={6 \over \pi^2 \rho^4} \left( {\rho^2 \over x^2 + \rho^2}
\right)^4 \;,
\label{instc}
\end{equation}
where $\rho$ is the size parameter. Zero temperature calculations
using the cooling method \cite{chn} show that after about 50 cooling
steps, the gauge fields are dominated by large, isolated instantons,
whose profiles agree well with that given by the continuum analytic
expression Eq.~(\ref{instc}) with $\rho \approx 0.3$~fm. We will
monitor the temperature dependence of the topological charge density
correlation as given by Eq.~(\ref{cprofile}).

At zero temperature, due to Euclidean symmetry the space-like and
time-like plaquettes $(P_x, P_t)$ have equal expectation values. At
finite temperature, however, the Euclidean symmetry is broken, and an
asymmetry between the space-like and time-like plaquettes develops at
the phase transition temperature proportional to the entropy $s$ of
the system \cite{engels}:
\widetext
\begin{equation}
Ts= \epsilon + P=4\beta \left[ 1-0.16675g^2 + O(g^4) \right]
(\langle P_t\rangle - \langle P_x\rangle ) \;,
\label{latent}
\end{equation}
where
\begin{equation}
\langle P_t\rangle \equiv {1 \over 3N_s^3 N_t} \sum _{n,i} {1 \over
3}{\rm Tr} U_{0,i} (x_n) \;,
\end{equation}
and
\begin{equation}
\langle P_x\rangle \equiv {1 \over 3N_s^3 N_t} \sum _{n,i<j}{1\over
3}{\rm Tr} U_{i,j} (x_n) \;.
\end{equation}
\narrowtext
Since the pressure $p$ is continuous across the phase transition, the
discontinuity of $\epsilon + p$ is the latent heat $\Delta \epsilon$.
Understanding the physics origin of the latent heat is important to
developing models for finite temperature hadronic matter. For
example, it was postulated in Ref.~\cite{shur2} that instanton liquid
reorganizes into molecules near $T_c$ and this process contributes to
a jump in the energy density at $T_c$. Here we calculate the latent
heat from both cooled and uncooled configurations so as to monitor
the contribution of the instantons to this quantity.

To study the quark distribution in various hadronic channels, we
calculate the Bethe-Salpeter amplitudes \cite{scor,bsa}:
\widetext
\begin{eqnarray}
\Psi^{\rm BS}_\pi (y) \equiv \int d{\vec x}\bigl\langle
\Omega \vert \bar{d} ({\vec x})
P\,e^{i\int\limits^{{\vec x}+{\vec y}}_{\vec x}
A ({\vec x'}) d{\vec x'}} \gamma_5 u( {\vec x}+{\vec y})
\vert \pi\bigr\rangle \
\label{pion}
\\
\Psi^{\rm BS}_\rho (y) \equiv \int d{\vec x}\bigl\langle
\Omega \vert \bar{d} ({\vec x})
P\,e^{i\int\limits^{{\vec x}+{\vec y}}_{\vec x}
A ({\vec x'}) d{\vec x'}} \gamma_2 u( {\vec x}+{\vec y})
\vert \pi\bigr\rangle \;.
\label{rho}
\end{eqnarray}
\narrowtext
Here, $\vert \Omega\rangle$ is the vacuum state, and $P$ indicates
path-ordering of the gauge fields $A$ put in to make the amplitudes
gauge invariant. We have used the natural choice of a straight line
joining the two quarks at ${\vec x}$ and ${\vec x}+{\vec y}$. When
$T$ is finite, the above amplitudes are extracted at large $z$
instead of the usual Euclidean time filtering. This results in a
projection into the lowest ${\it momentum}$ state, and therefore one
cannot interpret the resulting amplitudes as a ${\it wave function}$.
However, some insights into hadronic structure can be obtained by
comparing the QCD calculations to model results
\cite{scor,koch,schafer}. In particular, one can compare the
amplitudes for free quarks with those calculated at various
temperatures, to check whether at some temperature, hadronic matter
behaves like weakly interacting quark-gluon gas. Such a comparison
has been done in two recent papers \cite{scor}, with the surprising
result that the Bethe-Salpeter amplitudes seems to be quite
insensitive to changes in temperature up to $T=1.5 T_c$. In this
paper, we extend the previous calculations to higher temperature $T=2
T_c$, and we also calculate the amplitudes from cooled configurations
so that we can correlate any changes in the amplitudes with the
instanton content of the gauge fields.

\section{Cooling}
\label{seccool}

The method of lattice cooling has been extensively studied and
discussed in the literature \cite{htw}. The method can be applied to
finite temperature configurations without modifications. We used the
same procedure as in Ref.~\cite{chn}.

Briefly speaking, the cooling method smooths out the gauge fields
locally, by minimizing the action density. After some cooling steps
the short-range fluctuations in the gauge fields are suppressed,
leaving the long-range ``bumps'' more or less unchanged. In
particular, large isolated instantons, which are protected by
topology, dominate the gauge fields after many cooling steps, while
the smaller instantons annihilate with nearby anti-instantons. The
topological charge approaches a plateau corresponding to an integral
number of instantons when plotted against the number of cooling
steps. The topological susceptibility then becomes constant, since
the annihilation of instanton pairs does not change its value. Only
then can one extract an unambiguous value for the susceptibility
using the cooling method. In Fig.~1 we show the cooling history of
$\langle Q^2\rangle $ for several $N_t$. In contrast to
Ref.~\cite{dig}, we have no problem extracting $\langle Q^2\rangle $
and therefore the susceptibility even for $T>T_c$.

\begin{minipage}[t]{6truein}{
$$\epsfxsize=4.75truein\epsffile{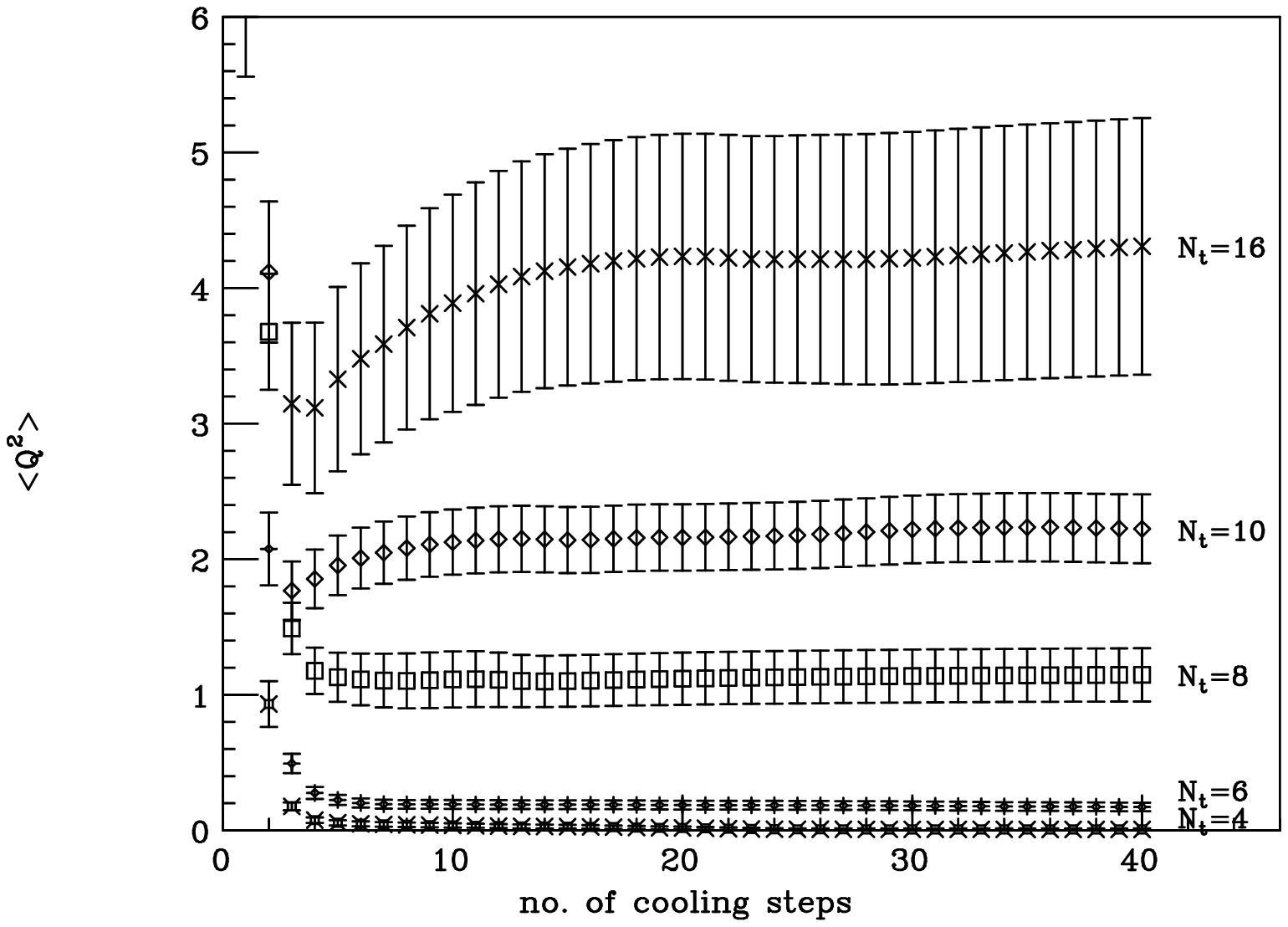}$$
\vbox{\footnotesize\baselineskip=12pt FIG. 1. The expectation value
of topological charge squared vs.\ number of cooling steps for
$N_t=4$, 6, 8, 10, and $16$. The quenched configurations are
generated using $\beta=6$ and $N_s=16$. In all cases, the values of
$\langle Q^2\rangle $ become unchanged after only about 15 cooling
steps. Same behavior was found for $N_t=12$ and $14$ as well, though
not shown here for clarity. The error bars reflect statistical
uncertainties in the configuration averaging only.}}
\vspace{.5cm}
\end{minipage}

After many cooling steps, the instantons are well separated in the
gauge fields, and the dilute gas approximation should become valid.
The probability distributions for topological charge is then given by
a convolution of Poissonian distributions \cite{htw}:
\begin{equation}
P(Q)=e^{-2m} m^{Q} \sum _{i=0}^{\infty} {m^{2i} \over i! (i+Q)!} \;,
\label{poisson}
\end{equation}
and the dilute gas value of $\langle Q^2\rangle $ is then given by
\begin{equation}
\langle Q^2\rangle=2m \;.
\end{equation}
In the case of an ideal Poissonian process, $m$ would be equal to the
number of instantons (anti-instantons). A comparison of the
distribution of topological charge at various $N_t$ with the dilute
gas approximation (dashed lines) is given in Fig.~2. Their agreement
gives a nontrivial check that our gauge fields are dominated by
dilute instanton ensembles after our cooling procedure, and the
topological susceptibility we extract from Fig.~1 reflects the
fluctuations in the topological charge in the ensembles.

\begin{minipage}[t]{6 truein}{
$$\epsfysize=4truein \epsffile{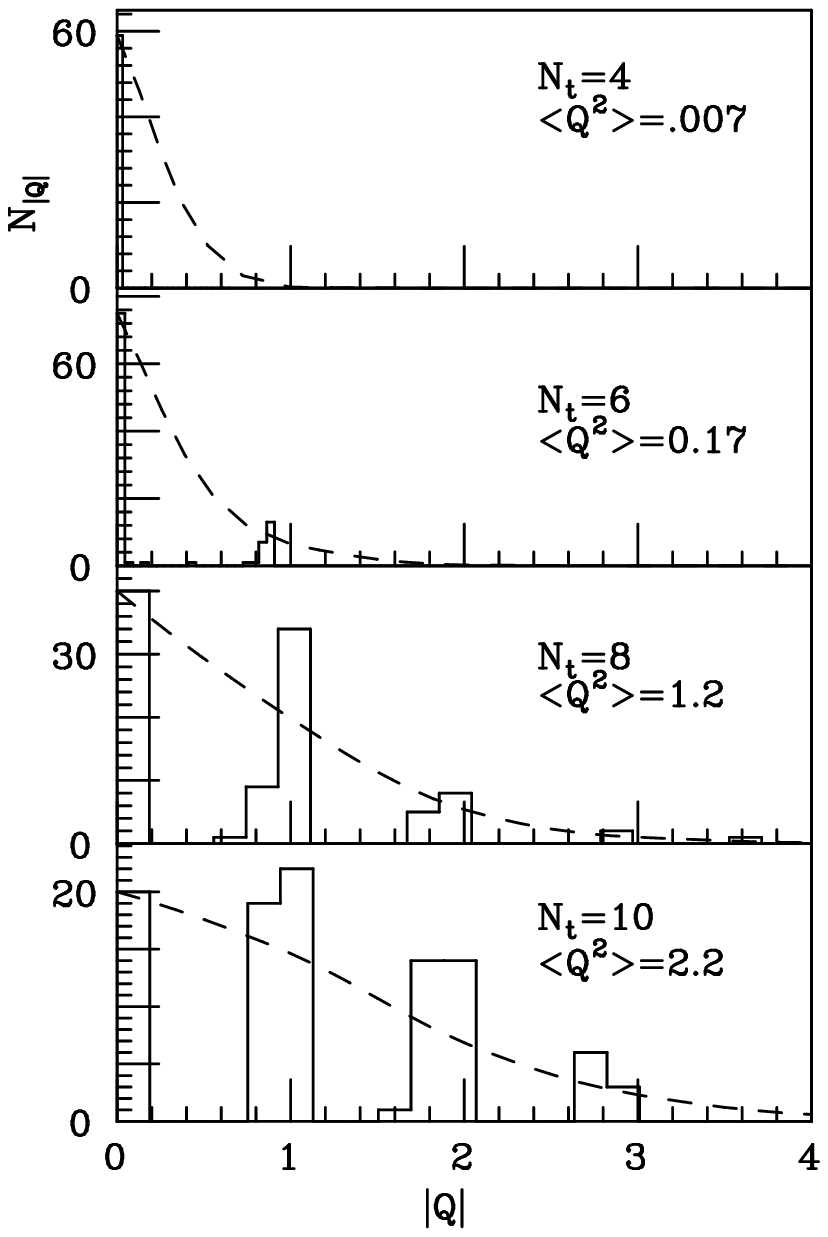}$$
\vbox{\footnotesize\baselineskip=12pt FIG. 2. Histograms showing the
distribution of topological charge in the configurations for $N_t=4$,
6, 8, and 10. The dashed lines show the distribution given by the
dilute gas approximation (see Sec.~III), using the $\langle
Q^2\rangle $ read from Fig.~1. Notice the clustering of data around
integral values of the topological charge.}}
\vspace{.5cm}
\end{minipage}

\section{Results}
\label{secres}
For the generation of the gauge field configurations we used the
Metropolis algorithm. We studied a lattice with spatial volume
$V_s=16^3$. Following Eq.~(\ref{temp}), we varied the temperature by
varying the number of time slices using values $N_t=4$, 6, 8, 10, 12,
14, and 16, respectively. The coupling constant is chosen to be
$\beta=6.0$ which corresponds to a physical lattice spacing $a\sim
0.1$~fm. Thus the calculation covers a temperature range between 125
and 500~MeV. For this value of $\beta$ the critical temperature is
$T_c \sim 250$~MeV. We generated 100 configurations each for $N_t=4$,
6, 8, 10, 12, and 40 configurations for $N_t= 14$ and $16$, which
were separated by 300 lattice sweeps. The configurations were then
cooled for 40 steps using the Cabibbo-Marinari heat bath algorithm in
the limit of infinite coupling strength $b=\infty$ as discussed in
Ref.~\cite{chn}. We verified the stability of our results by varying
the number of cooling steps.

\begin{minipage}[t]{6 truein}{
$$\epsfxsize=4.5truein \epsffile{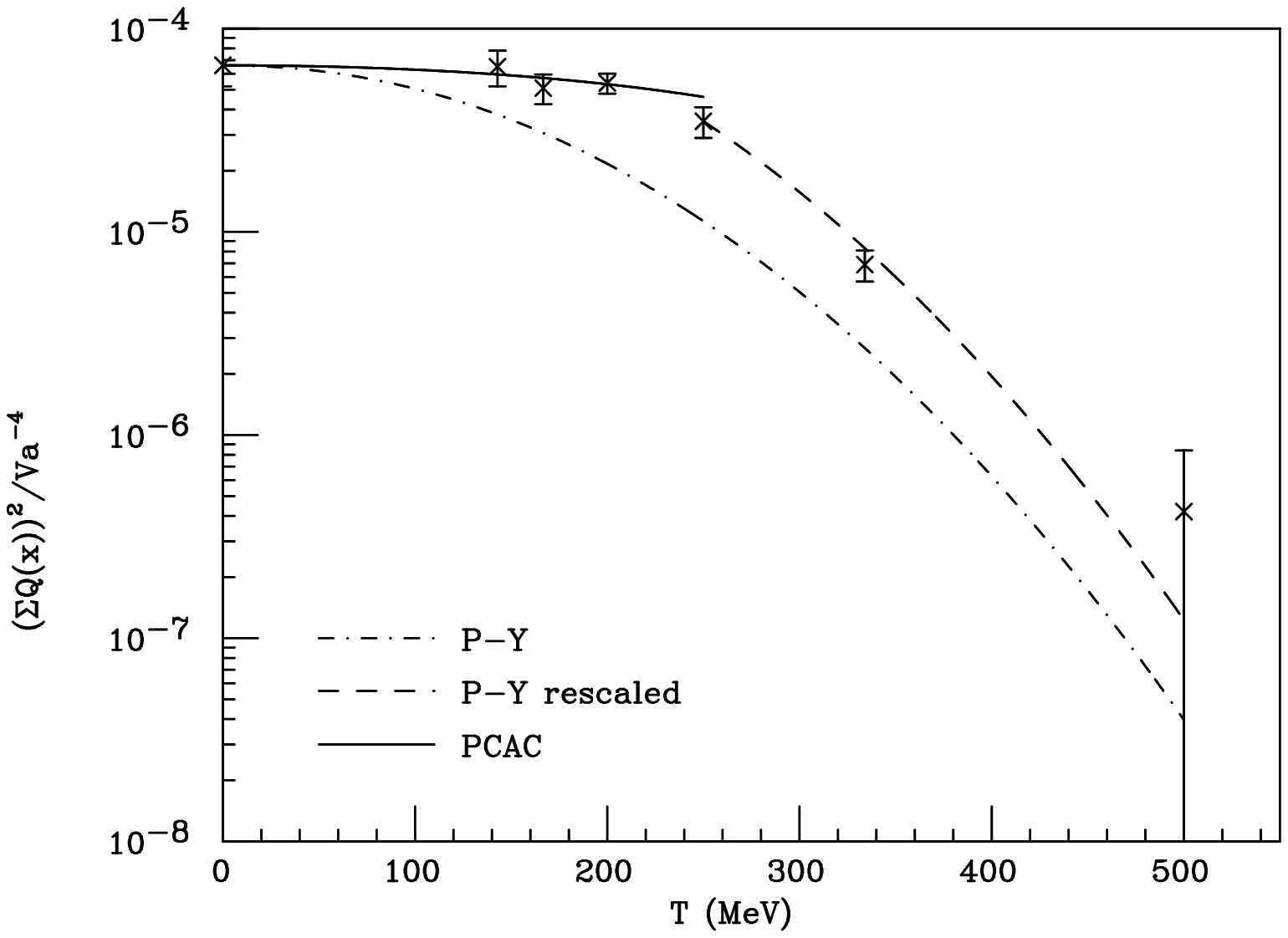}$$
\vbox{\footnotesize\baselineskip=12pt FIG. 3. Topological
susceptibility as a function of temperature. Lattice data, shown with
error bars, are compared with: 1) the PCAC expectation assuming soft
pion gas as the heat bath, taking $c=-1/6$ and $F_\pi=186$~MeV (see
Eq.~15), indicated by the solid line; 2) the Pisarski-Yaffe formula
(Eq.~16, using $\rho=0.26$~fm), derived assuming Debye-screening
mechanism of instanton suppression, indicated by the dotdashed line;
3) the Pisarski-Yaffe formula rescaled to match with the value of the
topological susceptibility at $T_c=250$~MeV, shown in dashed line.
The phenomenological value is also plotted at zero temperature.}}
\vspace{.5cm}
\end{minipage}

The numerical values for the topological susceptibility are shown in
Fig.~3. The plot also shows the approximate phenomenological value
for $\chi_t$ at zero temperature as estimated by Witten and Veneziano
\cite{Witten}
\begin{equation}
\chi_t=\frac{f_{\pi}^2}{4 N_f} (m_{\eta'}^2 - 2 m_K^2 + m_{\eta}^2)
\sim (180~{\rm MeV})^4 \;.
\end{equation}
As can be seen, the phenomenological value and the numerical result
for the $V=16^4$ lattice agree quite well. The temperature behavior
of $\chi_t$ exhibits a gentle decrease for $T<T_c$, turning into a
relatively	 sharp decay of the susceptibility around the
critical temperature $T_c$. At low temperature, $T\ll T_c$, one can
model the heat bath as a soft pion gas and apply PCAC methods
\cite{shur2.5} to obtain
\begin{equation}
\chi _t (T)=\chi _t (T=0) \left( 1+cT^2 / F_\pi^2 \right) \;,
\label{lowt}
\end{equation}
with $-1/6 \leq c \leq 1/6$. A very mild temperature dependence is
thus predicted for low temperature. We show in Fig.~3 the maximum
suppression of the topological susceptibility $(c=-1/6)$ by the solid
line, which agrees quite well with our data up to $T \approx
200$~MeV, though our statistics is not good enough to narrow down the
range of $a$. The agreement is probably accidental though, since our
quenched configurations should not reproduce the physics of a thermal
pion gas. The instanton density is suppressed by a factor of one half
when $T=T_c$, with some stronger suppression mechanism operative at
temperatures between $T= 200$~MeV and $T_c$. A rather sharp decrease
of the topological susceptibility continues above the phase
transition. At temperature $T\sim 334$~MeV topological effects are
strongly suppressed; $\chi_t$ is practically zero at $T\sim 500$~MeV.
The dotdashed curve in Fig.~3 shows the temperature dependence for
$\chi_t$ as predicted by Pisarski and Yaffe\cite{py} calculating the
Debye-screening suppression of large scale instantons:
\widetext
\begin{equation}
\chi_t (T)=\chi_t (T=0) \left( 1 + \lambda^2 /3 \right)^{3 \over 2}
\exp \left[ -2\lambda^2 -18 \alpha \left( 1+\gamma \lambda^{-3 \over
2}
\right)^{-8} \right] \;,
\label{pyf}
\end{equation}
\narrowtext
for quenched $SU(3)$ fields, with $\lambda \equiv \pi \rho T$,
$\alpha=0.01289764$, $\gamma=0.15858$, and $\rho=0.26$~fm is the size
parameter of instantons we obtained by fitting our $N_t=6$ data (see
discussion below), which differs slightly from a previous lattice
calculation at zero temperature \cite{chn}. The perturbative result
in Eq.~\ref{pyf} is supposed to be valid only at very high
temperature. In the temperature range we are considering, the
perturbative formula gives too large a suppression. The dashed curve
in Fig.~3 was computed by rescaling the Pisarski-Yaffe expression to
match with the susceptibility at $T_c=250$~MeV, which then seems to
be in agreement with the data, though more data is needed between
$T_c$ and $2T_c$ for a more meaningful comparison. The
phenomenological consequences of the restoration of $U_A(1)$, such as
changes in the $\eta'$ mass, should be an exciting subject for
experimental works at AGS, CERN, and RHIC.

\begin{minipage}[t]{6 truein}{
$$\epsfxsize=4truein \epsffile{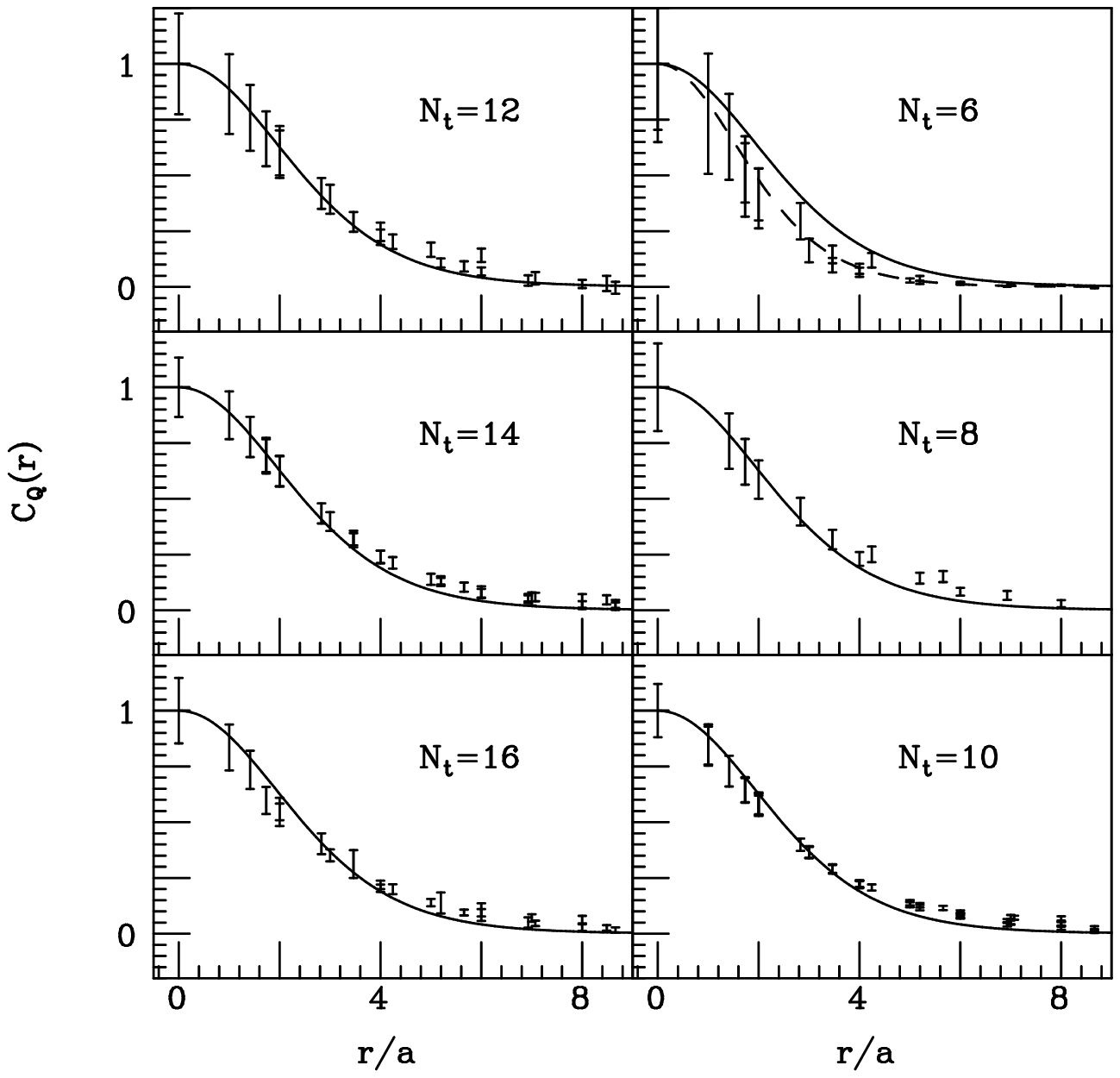}$$
\vbox{\footnotesize\baselineskip=12pt FIG. 4. Topological charge
correlation function, calculated for $N_t=6$, 8, 10, 12, 14, and 16.
The solid lines show the correlation function calculated with the
continuum instanton charge distribution assuming a size parameter
$\rho=3.3 a$. The dashed line shown at $N_t=6$ is the continuum
instanton correlation function with $\rho=2.6 a$.}}
\vspace{.5cm}
\end{minipage}

In Fig.~4 we show the correlation function $C_Q(x)$,
Eq.~(\ref{cprofile}), with $x$ taken in the spatial direction. Six
different temperatures are shown. For comparison the plots include
the correlation function calculated with an isolated instanton with
size parameter $\rho=3.3a \sim 0.33$~fm, which fits the data
reasonably well for all temperatures up to $T_c$, at $N_t=8$. At
$T\approx 334$~MeV, or $N_t=6$, the instanton profile with $\rho=2.6
a \sim 0.26$~fm gives a better fit to the data. In Fig.~5, we show
the cooling dependence of the topological charge correlation function
for $N_t=6$. The shape of the correlation is very stable between 20
and 100 cooling steps, and the smaller size parameter $\rho=0.26$~fm
is clearly favored by the data over the larger one ($\rho=0.33$~fm)
preferred at lower temperatures. As can be seen, the temperature
dependence is quite weak below $T_c$, but our data suggests a rather
sudden change in $\rho$ across the phase transition. This may signal
major reorganization in the gluon fields above $T_c$. For $N_t=4$,
the topological charge correlation function changes continuously as a
function of cooling steps, and therefore we could not extract a
meaningful shape at that temperature. The correlation function
becomes flat at small $N_t$ when $x$ is taken in the temporal
direction because of the periodic boundary condition.

\begin{minipage}[t]{6 truein}{
$$\epsfxsize=4.75truein \epsffile{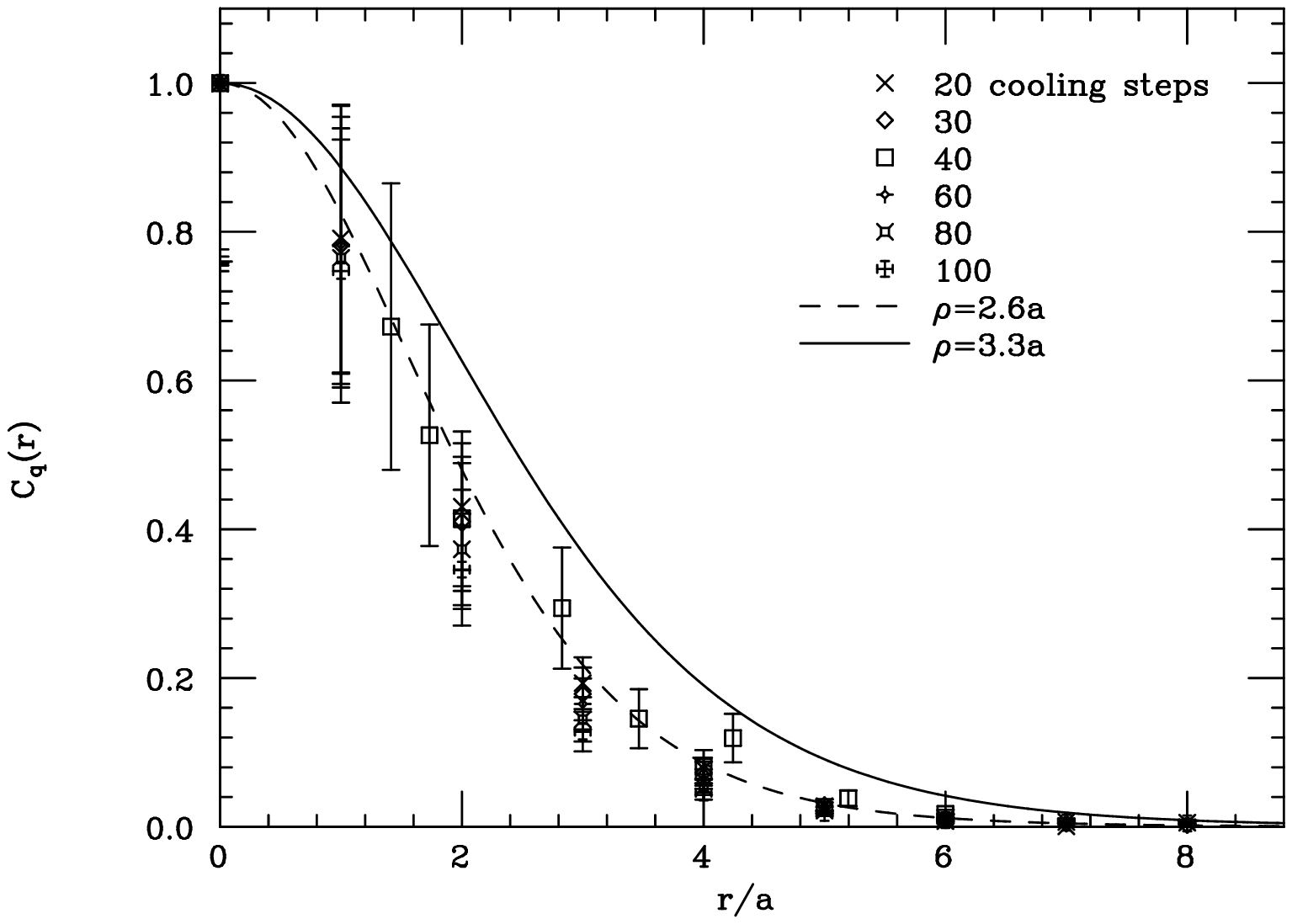}$$
\vbox{\footnotesize\baselineskip=12pt FIG. 5. Topological charge
correlation function for $N_t=6$, at different cooling steps. The
solid and dashed lines are the same as in Fig.~4.}}
\vspace{.5cm}
\end{minipage}

\begin{minipage}[t]{6 truein}{
$$\epsfxsize=4truein \epsffile{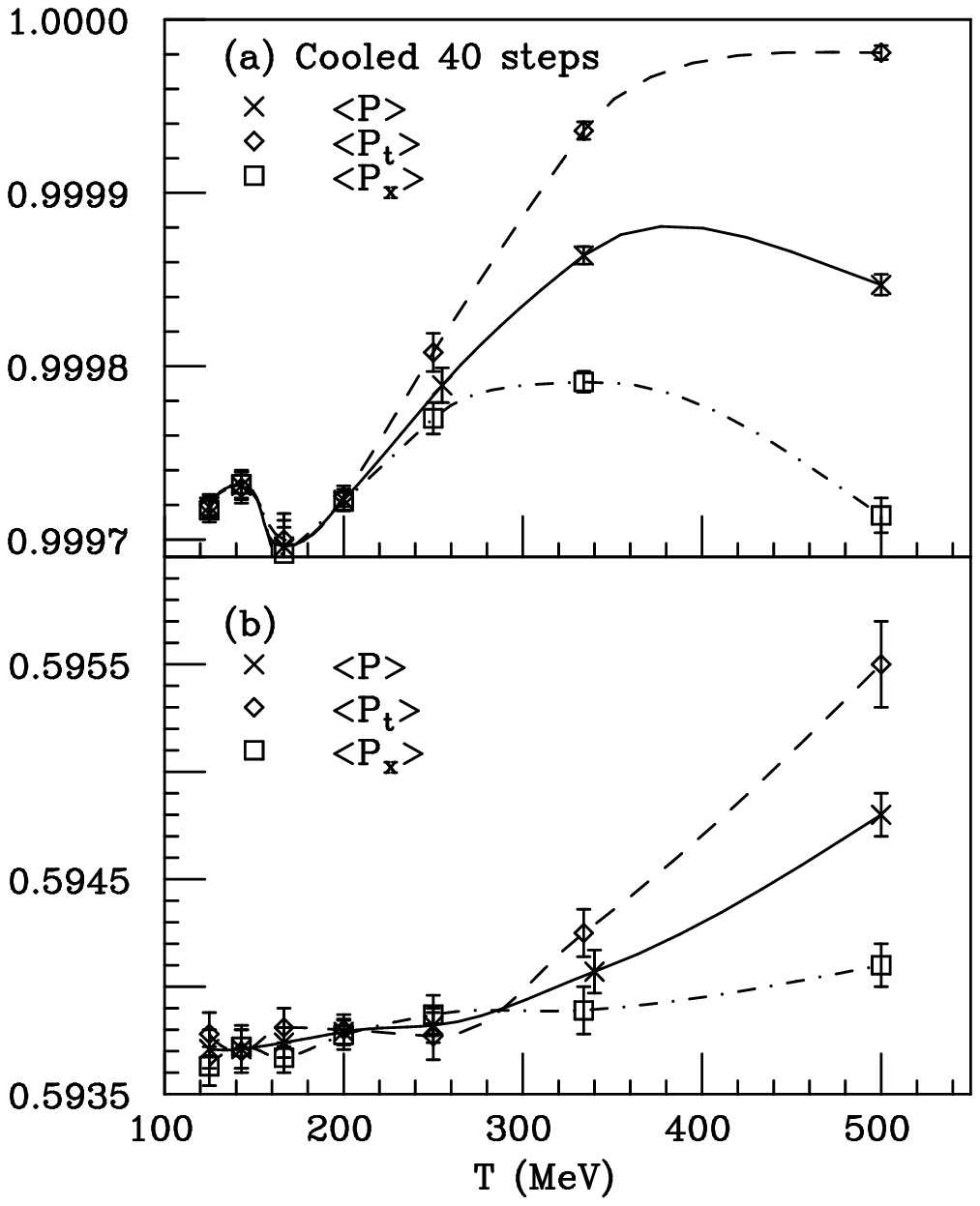}$$
\vbox{\footnotesize\baselineskip=12pt FIG. 6. Plaquette expectation
values as a function of temperature for (a) cooled and (b) uncooled
configurations. The space-like plaquette expectation values $\langle
P_x\rangle$ (squares) are compared to the time-like ones $\langle
P_t\rangle$ (diamonds) and the overall averages $\langle P\rangle$
(crosses). The lines are drawn to guide the eyes only.}}
\vspace{.5cm}
\end{minipage}

We monitor the contribution of the large, isolated instantons to the
latent heat by calculating the splittings of the spatial and temporal
plaquettes for both cooled and uncooled configurations. The numerical
values of the plaquettes are plotted in Fig.~6 (a) and (b), where a
clear splitting between the spatial and temporal plaquettes, $\Delta
P \equiv \langle P_t\rangle - \langle P_x\rangle $, can be seen for
temperatures above the phase transition. The splitting $\Delta P$,
which is proportional to the latent heat, is clearly a function of
cooling, and it approaches zero in the limit of infinite cooling
steps. An estimate of the fraction of latent heat contributed by
instantons, $f$, is given by
\begin{equation}
f \approx {\Delta P _{\rm cooled} \over
\Delta P _{\rm uncooled}} \;.
\label{lat}
\end{equation}
In Fig.~7 we show $f$ as a function of cooling steps for both $N_t=4$
and 6. In both cases, $f$ drops very quickly in the first 40 cooling
steps, showing that most of the splitting is due to short range
fluctuations that got eliminated by cooling. After about 50 cooling
steps, the $N_t=4$ splitting levels off to a plateau, where its value
represents an upper bound to the contribution of the instantons at
$T=500$~MeV. The splitting at $N_t=6$ has not reached a plateau yet
even at 100 cooling steps; we quote just the fraction $f$ at 100th
cooling step as the upper bound at $T=334$~MeV. Our results for $f$
are $0.15 \pm 0.04$ at $T=334$~MeV and $0.056 \pm 0.004$ at
$T=500$~MeV.

\begin{minipage}[t]{6 truein}{
$$\epsfxsize=4.75truein \epsffile{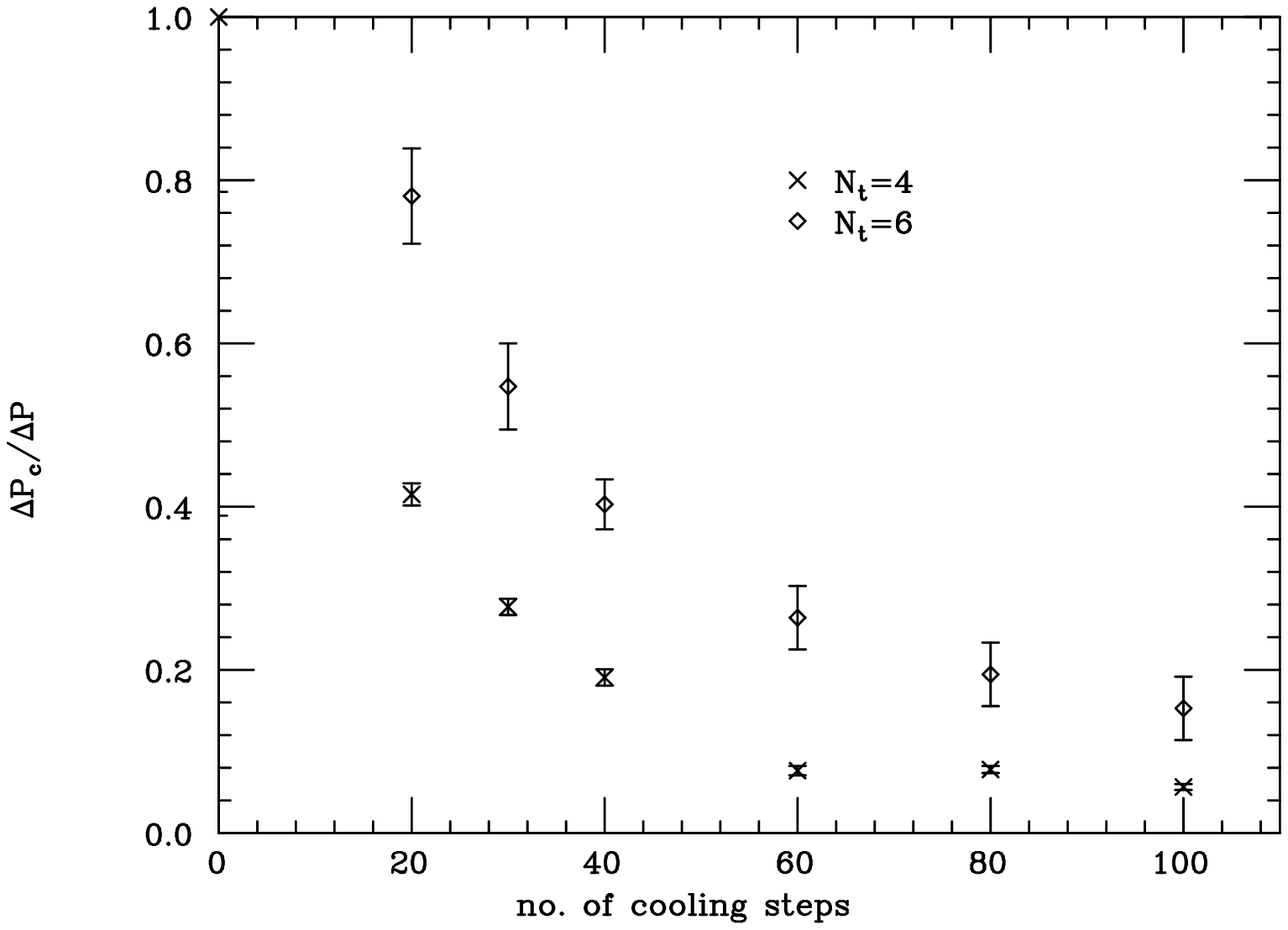}$$
\vbox{\footnotesize\baselineskip=12pt FIG. 7. Plaquette splittings,
$\Delta P_c=\langle P_t\rangle - \langle P_x\rangle$, normalized to
their uncooled values, as a function of cooling steps. $N_t=4$ data
are shown as crosses, and $N_t=6$ data as diamonds. Values at 100th
cooling step represent upper bounds to the fraction of latent heat
contributed by the instantons.}}
\vspace{.5cm}
\end{minipage}

Finally, we computed the screening wave functions for the $\pi$ and
$\rho$ mesons as explained in Sec.~\ref{secbasic}. We used a
conjugate gradient procedure to calculate the Greens functions for
Wilson fermions with $\kappa=0.1541$, which corresponds to a quark
mass $m\sim 100$~MeV. The average of a set of ten wave functions for
$N_t=10, 6, 4$ corresponding to $T=200, 334, 500$~MeV are shown in
Figs.~8 and 9. As already observed in \cite{scor} the screening wave
functions change only slightly as a function of temperature across
$T_c$. The spatial extension of the wave functions shrinks as
temperature increases, in agreement with theoretical arguments
discussed in Refs.~\cite{scor,koch}. Here the main idea is that in
the deconfined phase, the quark mass for propagation along the
spatial axis is essentially given by the lowest Matsubara frequency
$m \sim \pi T$. Strong correlations resulting from unscreened
magnetic interactions generate a bound state between quarks
propagating along a spatial axis even in the deconfined phase. The
effective quark mass increases at high temperatures, thus decreasing
the characteristic size of the state. However, the weak dependence of
the wave functions is surprising in view of the factor of 2.5
increase in the effective quark mass from $N_t=10$ to $4$. It is
probable that the present temperature range is not high enough for
the above picture based on dimensional reduction to be accurate.

\begin{minipage}[t]{6 truein}{
$$\epsfxsize=4.75truein \epsffile{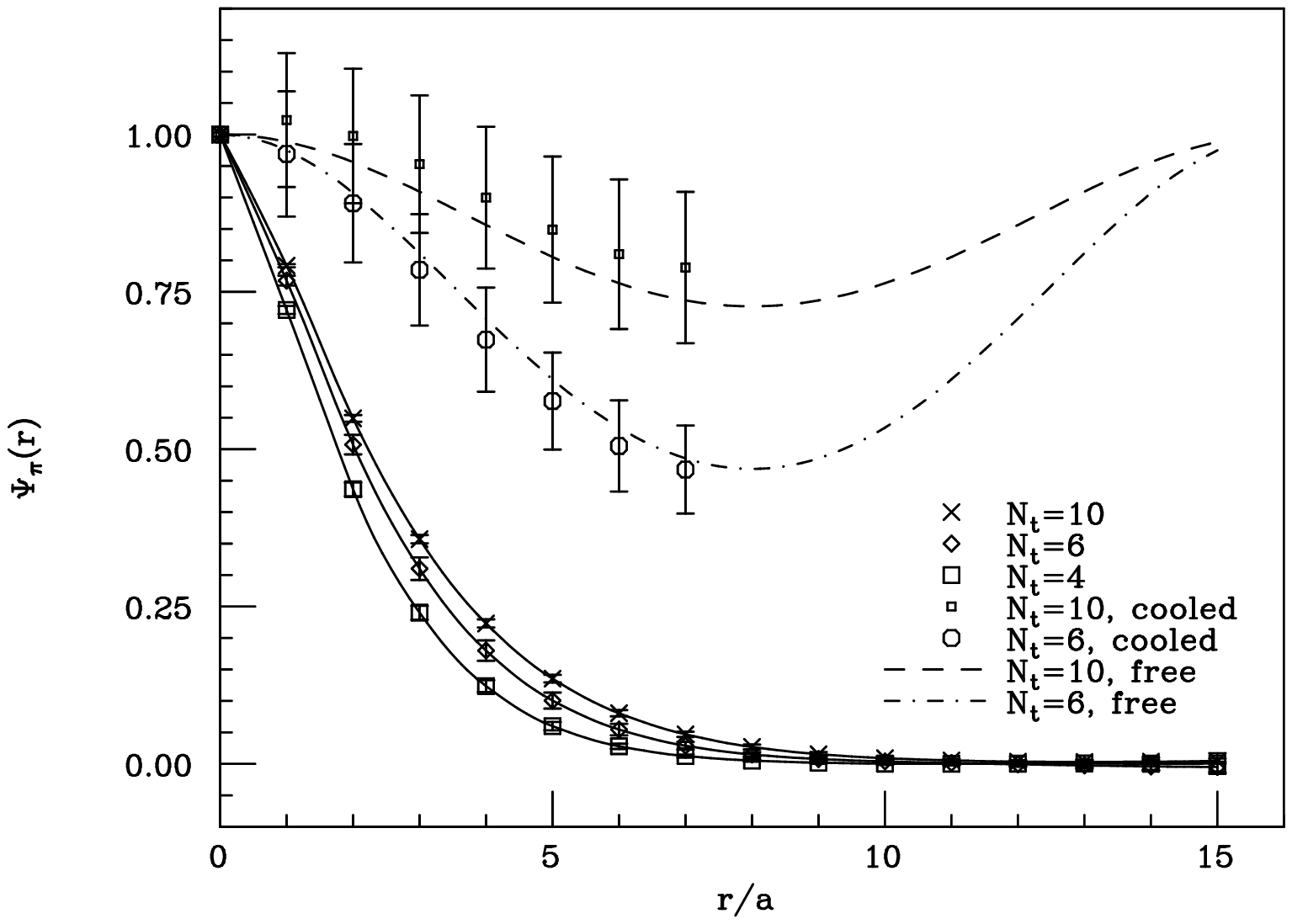}$$
\vbox{\footnotesize\baselineskip=12pt FIG. 8. Screening wave
functions of the $\pi$ meson, for $N_t= 4$ (squares), 6 (diamonds),
and 10 (crosses), corresponding to $T=500$, 334, 200~MeV,
respectively. wave functions calculated with cooled configurations
and symmetrized to take into account of the periodic boundary
conditions are also shown for $N_t=6$ (circles) and 10 (small
squares), which agree well with free wave functions (dashed and solid
lines).}}
\vspace{.5cm}
\end{minipage}

\begin{minipage}[t]{6 truein}{
$$\epsfxsize=4.75truein \epsffile{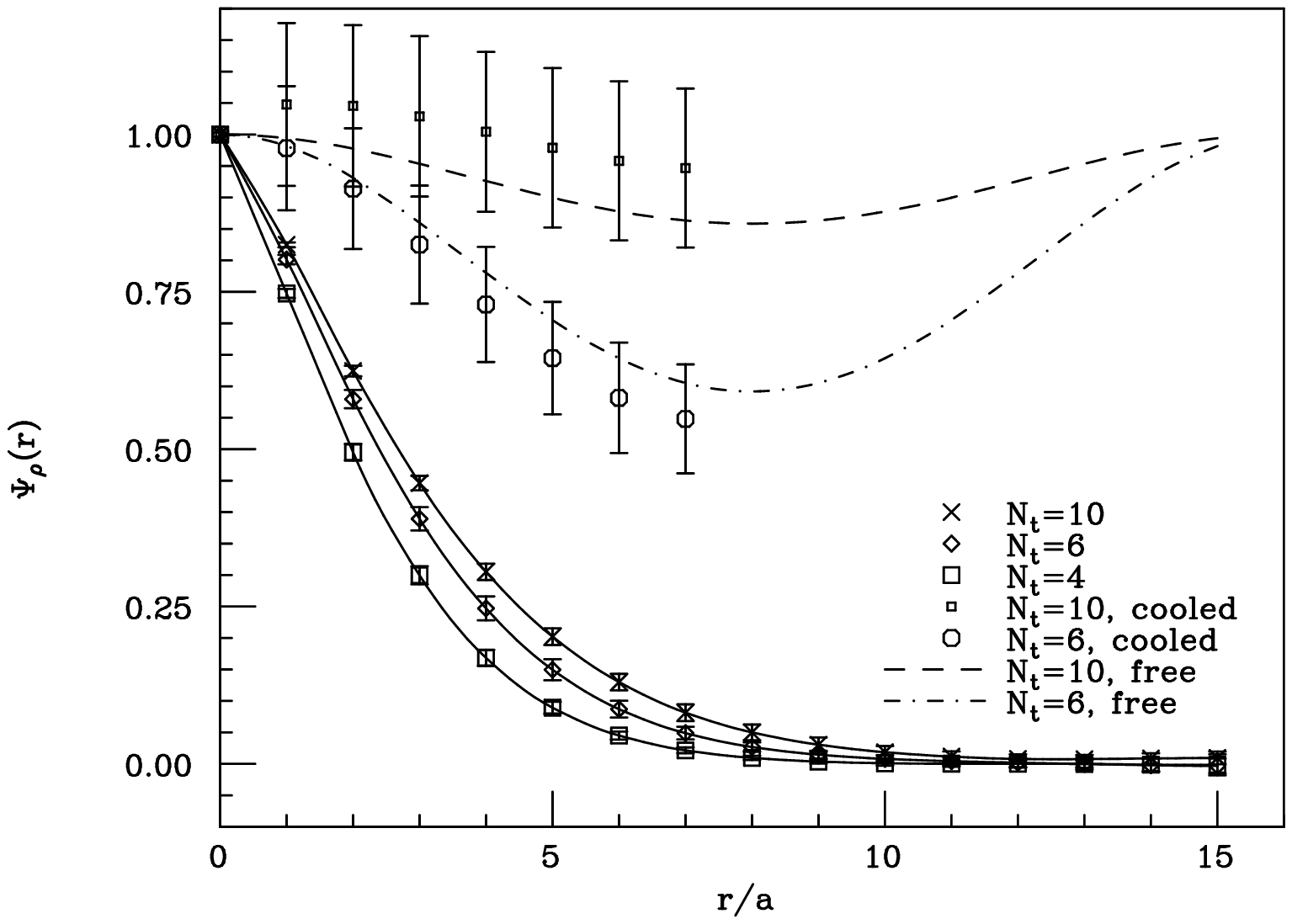}$$
\vbox{\footnotesize\baselineskip=12pt FIG. 9. Screening wave
functions of the $\rho$ meson, see Fig.~8.}}
\vspace{.5cm}
\end{minipage}

\newpage
As in the case of zero temperature, cooling effectively removed short
range fluctuations in the gauge fields responsible for quark
confinement as well as the Coulomb force. As a result, the wave
functions calculated from the cooled configurations are distinctly
broader, with the slope at short distances strongly reduced
\cite{chn}. To study the contribution of instantons to the screening
wave functions at such high temperatures, we compare the cooled wave
functions with the uncooled ones as well as the free wave functions.
As shown in Figs.~8 and 9, for both $\pi$ and $\rho$ at $T=200$ and
$500$~MeV, after symmetrized to take into account the periodic
boundary conditions, the cooled wave functions are indistinguishable
from the free ones. Our results imply that the large scale instantons
do not play an important role in the screening wave functions in this
range of the temperature. This is possibly a result of both the
suppression of instanton density and the increase of the effective
quark mass by the Matsubara frequency. Instead, some non-trivial
interactions are operative, which correlate the effective massive
quarks. It is important to understand this mechanism in order to
develop an accurate phenomenology of hadronic physics at temperatures
around $T_c$, just where the current and future heavy-ion experiments
probe.

\section{Summary}
\label{seccon}
We discussed the content and the role of instantons in quenched QCD
matter at finite temperature. Specifically, we calculated the density
of instantons and found that it is suppressed as temperature
increases. While the mild dependence of $\chi_t$ can be understood by
a PCAC method assuming a soft pion gas as the heat bath, the high
temperature ($T>T_c$) behavior deviates significantly from the
perturbative formula expressing the Debye-screening suppression of
instantons. Our results indicate that the restoration of $U_A(1)$
symmetry is a continuous process starting below $T_c$; at $T_c$,
about half of the axial anomalies are already removed. Some
non-perturbative mechanism enhances the rate of $U_A(1)$-restoration
in the temperature range $T_c \leq T \leq 2T_c$. The topological
charge correlation function agrees well with the continuum instanton
profile, though our data suggests a rapid change in the size
parameter from $\rho=0.33$~fm at and below 250~MeV to $0.26$~fm at
334~MeV. By comparing the entropy in the cooled and uncooled gauge
fields, we obtain upper bounds --- of about 5\% at $2T_c$ and 15\% at
$1.5T_c$ --- on the instanton contribution to the latent heat. The
screening wave functions of $\pi$ and $\rho$ mesons show a weak
temperature dependence for temperatures as high as $2T_c$. This
points to strong correlations between quarks even in the deconfined
phase. In contrast to the zero temperature situation, the screening
wave functions at temperatures above $T \approx 200$~MeV are not
dominated by the instantons.

It will be interesting to compare these results with unquenched
calculations, as it is quite possible that the reorganization of the
gauge fields near the phase transition is sensitive to the presence
of dynamical quarks. Much more theoretical work is needed to
illuminate the complex and drastic changes of hadronic physics near
the phase transition.
\vspace{.5cm}

We thank Suzhou Huang and Edward Shuryak for stimulating discussions.
We thank the San Diego Supercomputer Center as well as the National
Energy Research Supercomputer Center for providing Cray-C90 computer
resources. This research is supported in part by the National Science
Foundation, Grant Nos. PHY90-13248 and PHY94-12818, at Caltech and
the U.S. Department of Energy, Grant No. DE-FG02-87ER40365, at
Indiana University.

\end{document}